# Synthetic aperture laser optical feedback imaging using a translational scanning with galvanometric mirrors

Wilfried Glastre,* Olivier Jacquin, Olivier Hugon, Hugues Guillet de Chatellus, and Eric Lacot

*Centre National de la Recherche Scientifique / Université de Grenoble 1,*

*Laboratoire Interdisciplinaire de Physique, UMR 5588,*

*Grenoble, F- 38041, France*

*\*Corresponding author: wilfried.glastre@ujf-grenoble.fr*

In this paper we present an experimental setup based on Laser Optical Feedback Imaging (LOFI) and on Synthetic Aperture (SA) with translational scanning by galvanometric mirrors for the purpose of making deep and resolved images through scattering media. We provide real 2D optical synthetic-aperture image of a fixed scattering target with a moving aperture and an isotropic resolution. We demonstrate theoretically and experimentally that we can keep microscope resolution beyond the working distance. A photometric balance is made and we show that the number of photons participating in the final image decreases with the square of the reconstruction distance. This degradation is partially compensated by the high sensitivity of LOFI.

OCIS codes: 070.0070, 090.0090, 110.0110, 180.0180.

## 1) Introduction

Making images with a good in-depth resolution through scattering media is a major issue, limited by a double problematic: first the scattering medium generally attenuates strongly the ballistic photons which enable to obtain resolved images and second, the wavefront is highly perturbed by scattered photons, degrading the quality of the resolved image. Several ways to overcome these problems have been proposed among which we can distinguish two main families. The first one uses pre-compensation of the wavefront before propagation, to improve the resolution. This technique is used successfully both with optics or acoustic modality [1,2,3], but it requires an *a priori* knowledge of the medium. The second one selects ballistic photons while rejecting multi-diffused parasitic photons: Optical Coherence Tomography (OCT) [4] and confocal microscopy associated [5] or not [6] to non linear effects belong to this family as well as tomographic diffractive microscopy [7]. Our Laser Optical Feedback Imaging (LOFI) setup, based on optical reinjection in the laser cavity [8,9,10] belongs to this second family. LOFI has the advantage of providing a self-aligned and very sensitive optical system limited by shot noise [11,12] whatever the detector noise is. It is a very simple (no alignment needed and transportable system) setup compared to many other interferometer. Losses in ballistic photons are compensated by this high sensitivity due to LOFI while multi-diffused photons are rejected by the confocal intrinsic feature of this technique. Furthermore it gives access to both amplitude and phase of the retrodiffused optical electric field.

In order to solve the issue of making in-depth images, we propose a new configuration based on synthetic aperture by translational scanning with galvanometric mirrors. The technique was introduced first in Synthetic Aperture Radar imaging (SAR) [13,14] to overcome the fact that no large portable aperture component exists for radio waves. Synthetic Aperture consists in scanning the target with a diverging beam while recording amplitude and phase informations (accessible thanks to LOFI in our case) on the movement of the laser spot with respect to the target. It enables to realize a numerical focusing to recover a good resolution. In the optical field, it has first been applied to optical wavelengths with $CO_2$ [15] and Nd:YAG microchip laser sources [16,17,18] in what is called Synthetic Aperture Laser (SAL). In all these previous pieces of work (SAR and SAL), the scanning was made only in one direction. The recovery of image resolution by synthetic aperture operation was performed in one direction (the scanning direction) whereas in the other direction only telemetry (a chirped signal is used instead of a monochromatic one, the frequency of the beating between the reinjected photons and the emitted signal depends on the round-trip length) is used to improve the resolution. Moreover at the beginning of SAL, the target itself was moved while the laser source was fixed. A setup presenting the advantage in terms of vibrations limitations and measurement speed, to have a fixed object and a scanning laser have been proposed in 2006 [19]. However, it has the drawback to be quite complex and as we said before, presents anisotropic resolution due to 1D scanning. In our case, by using two dimensional scanning with galvanometric mirrors, we are able to recover an isotropic resolution with a complete 2D scanning. Here we demonstrate what we believe to be the first 2D optical synthetic-aperture image of a fixed, scattering target with a moving aperture and an isotropic resolution. This work is a continuation of [20,21,22] where a galvanometric rotation scanning of a fixed object was performed with LOFI. The problem was that the object was scanned angularly, implying a degradation of the resolution with the reconstruction distance. The setup we propose here is based on a translational scanning of the object and we show that it implies a conservation of the resolution whatever the reconstruction distance is.

In a first part we present our experimental setup and remind the principles of LOFI; in particular the translational laser scanning by galvanometric mirrors is introduced. We then present in a second part a complete study of synthetic aperture operation and show that we can keep microscope resolution beyond the working distance. To conclude, in a third and last part dedicated to photometric performances of the setup, we show that the final image quality degrades proportionally to the square of the distance of numerical refocusing. This drawback is partially compensated by the high sensitivity of LOFI.

## 2) Reminder on LOFI and presentation of the experimental setup

*Experimental setup*

Figure 1 shows a description of the LOFI [8,9] experimental setup. The laser is a cw Nd:YVO$_4$ microchip emitting about 85 mW power at λ = 1064 nm. This laser has a relaxation frequency near $F_R \approx 2$ MHz. On its first pass, the laser beam is frequency shifted by a frequency $F_e/2$ where $F_e$ is close to the relaxation frequency of the laser ($F_R \approx F_e$), and then sent to the bidimensional target by means of two rotating mirrors, respectively called $M_x$ and $M_y$. The first one allows scanning of the target in the horizontal direction (x direction) and the second one in the vertical direction (y direction). The angular orientations of the galvanometric mirrors are given by the angles $α_x$ and $α_y$, respectively. The beam diffracted and/or scattered by the target is then reinjected inside the laser cavity after a second pass in the galvanometric scanner and the frequency shifter. The total frequency shift undergone by the photons reinjected in the laser cavity is therefore $F_e$ which results in triggering relaxation oscillations of the microlaser and in amplifying the sensitivity of the device to the reinjected photons. A small fraction of the output beam of the microchip laser is sent to a photodiode. The delivered voltage is analyzed by a lock-in amplifier at the demodulation frequency $F_e$, which gives the LOFI signal (i.e. the amplitude and the phase of the electric field of the backscattered light). Experimentally, the LOFI images (amplitude and phase) are obtained pixel by pixel (i.e., point by point, line after line) by full 2D galvanometric scanning ($α_x$, $α_y$). We must now consider two possibilities:

- "Conventional" LOFI (Figure 1 and Figure 2 with $L = 0$) where we scan the object with a focused beam. We can get an amplitude [8,9] $|h(α_X,α_Y)|$ or phase [23,24] image $Φ_S(α_X,α_Y)$.

- Synthetic Aperture (SA) imaging LOFI [20,21,22] corresponding to imaging with a defocused beam (Figure 1 and Figure 2 with $L ≠ 0$). This raw complex image $h(α_X,α_Y)$ must be filtered to realize a numerical post focusing. It has the advantage, as we will see to make images beyond the working distance of the lens.

In the following, whatever the target position is, we index all parameters related to an image without any post processing with "R" (Raw) and those associated with a numerical refocusing with "SA" (Synthetic Aperture).

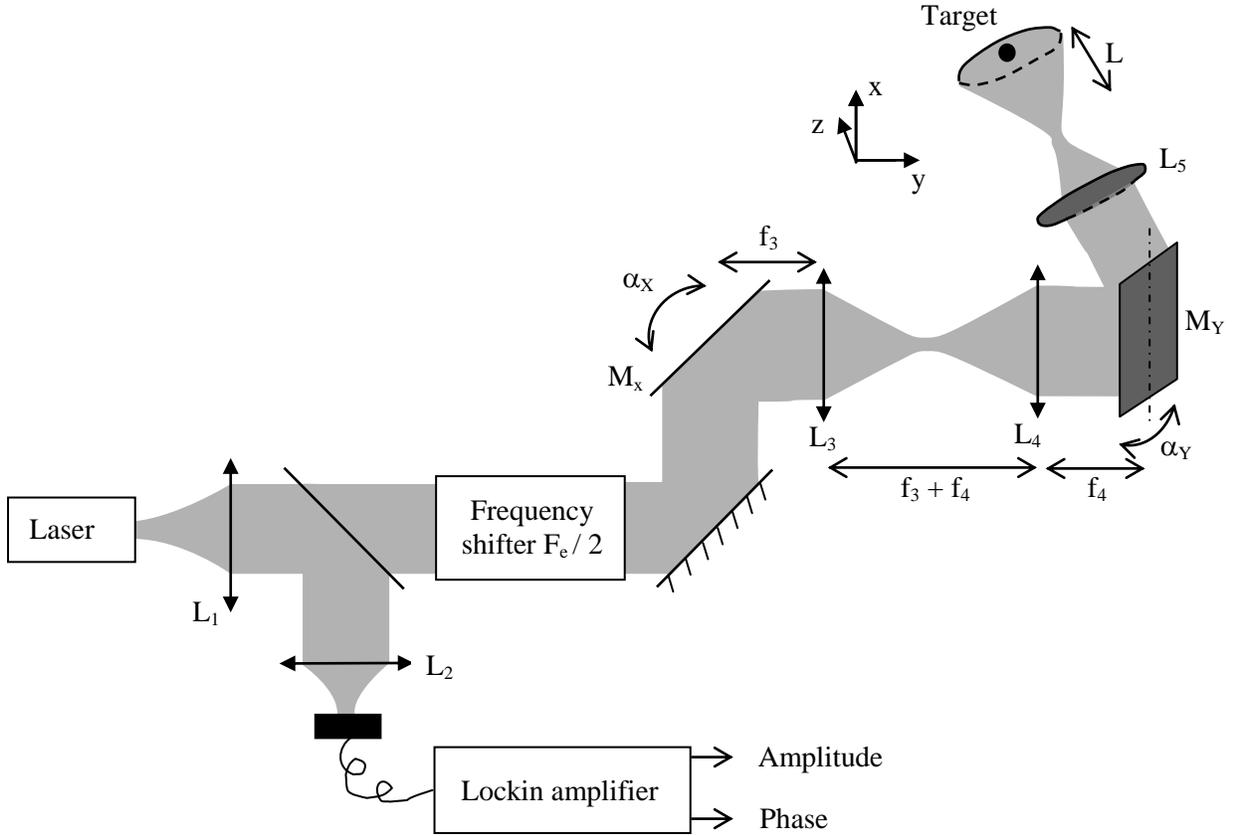

**Figure 1 : Experimental setup of the synthetic aperture LOFI-based imaging system. The laser is a cw Nd:YVO$_4$ microchip collimated by lens L$_1$. A beam splitter sends 10% of the beam on a photodiode connected to a lockin amplifier which give access to the amplitude and phase of the signal. The frequency shifter is made of two acousto-optic modulators which diffract respectively in orders 1 and -1 and gives a net frequency shift of F$_e$ / 2 = 1.5 MHz. X-Y plane is scanned (see Figure 2) by galvanometric mirrors M$_X$ (scan in the X direction) and M$_Y$ (scan in the Y direction) conjugated by a telescope made by lenses L$_3$ and L$_4$. f$_3$, f$_4$ and f$_5$ are the focal lengths of lenses L$_3$, L$_4$ and L$_5$. α$_X$ and α$_Y$ are the angular positions of galvanometric mirrors M$_X$ and M$_Y$.**

## *Experimental scanning*

In order to get 2D image we need to have a relative translation between the target and the laser beam. In our setup we choose to keep a fixed target and to use galvanometric mirrors. The main advantage of this approach is to avoid parasitic vibrations and to benefit from the scanning speed of the galvanometric mirrors. Synthetic Aperture using galvanometric mirrors has previously been presented in [20,21,22]. The object was scanned angularly by the two galvanometric mirrors and we showed that it was possible to obtain a resolved image from a raw defocus image after numerical refocusing. The main drawback was that the final synthetic resolution was degrading linearly with working distance (distance between the mirrors and the target) [20,21]. To avoid that problem we choose here a translational (instead of angular) scanning of the target and we will show in following parts that it implies a constant resolution independently from the working distance. In our scanning setup the two galvanometric mirrors ($M_X$ and $M_Y$) are conjugated by a telescope formed by lens $L_3$ and $L_4$ having the same focal length $f_3 = f_4 = 50$ mm leading to a magnification of 1 (see Figure 1); we call $M_X$' the image of $M_X$ through the telescope. To achieve a translational scanning in X and Y directions, the mirrors $M_X$' and $M_Y$ are placed in the focal object plane of lens $L_5$ (see Figure 2). Lens $L_5$ (focal length $f_5 = 25$ mm here) converts the angular movement of the collimated laser beam after mirrors $M_X$' and $M_Y$ into a movement of translation and focuses the beam with a waist radius $r \approx 13$ µm in the focal image plane of $L_5$. It is important to precise that we are scanning with low angular amplitude (around one hundred of miliradian corresponding to few mm of spatial field for $f_5$' = 25 mm) for the two galvanometric mirrors as a result we can neglect aberrations that telescope and lens $L_5$ could add. This result in a planarity of the displacement of the beam waist with an interferometric precision (important for Synthetic Aperture processing); this have been experimentally checked by scanning a plane mirror, we got white light fringe. To further simplify our problem we make in the following, the approximation of paraxial and Gaussian optical rays which seems reasonable considering our experimental values (waist r = 13.5 µm for λ = 1064 nm). The relations between the angular position ($\alpha_X, \alpha_Y$) of the mirrors $M_X$ and $M_Y$ and the position (x,y) of the laser in the focal image plane are then given by:

$$x = 2 f_5 \alpha_x$$
$$y = 2 f_5 \alpha_y \quad (1)$$

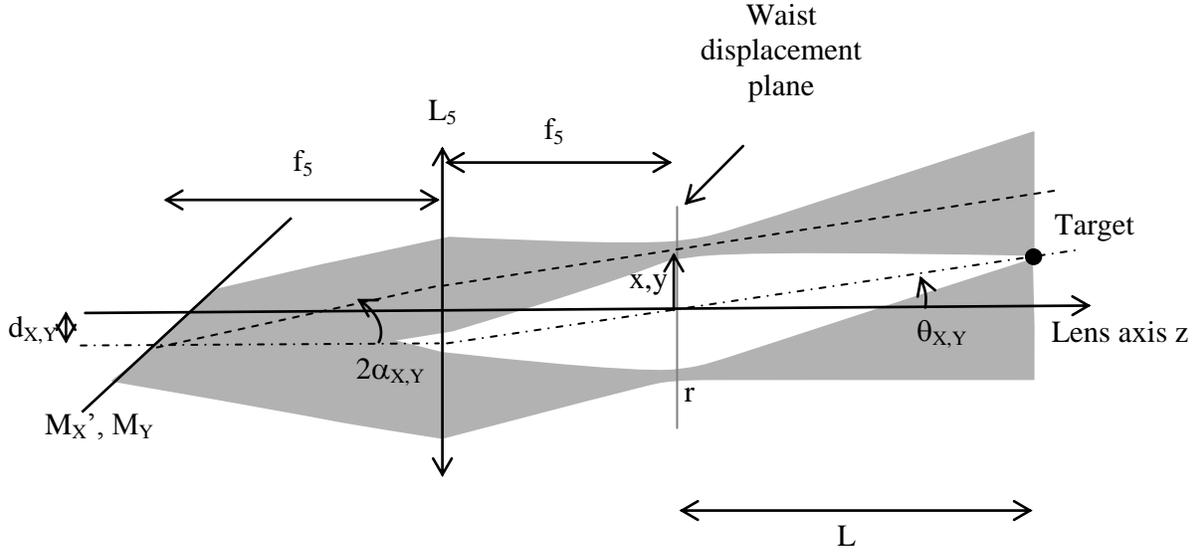

**Figure 2 :** Scanning mechanism of the setup of Figure 1 after lens $L_4$. The collimated laser beam is focused by the final imaging lens $L_5$ in its focal image plane with a waist radius r. Mirror $M_X$' (image of $M_X$ through telescope, see Figure 1) and $M_Y$ are in the focal object plane of imaging lens $L_5$, which implies a displacement of the beam waist in the image focal plane of $L_5$. The laser beam presents a slight misalignment with $L_5$ optical axis and rotation axis of galvanometric mirror of $d_X$ in X direction and $d_Y$ in the Y direction. $\alpha_X$ and $\alpha_Y$ are the angular position of galvanometric mirrors $M_X$' and $M_Y$. When $\alpha_X = \alpha_Y = 0$, the center of the beam is passing through the target (convention). x and y are the position of the waist in the focal image plane of $L_5$. $\theta_X$ and $\theta_Y$ are the angles of the center of the beam relative to the direction normal to the focal image plane of $L_5$ ($\neq 0$ due to $d_X$ and $d_Y$). The target at a position ($L\theta_X$, $L\theta_Y$) is scanned by a defocused beam at a distance L from the beam waist.

Due to mechanical imprecision, there can be a slight misalignment $d_X$ and $d_Y$ between the laser beam axis and mirrors $M_X$' and $M_Y$ rotation axis. As a result the angles $\theta_X$ and $\theta_Y$ between laser beam axis and the Z direction ( Figure 2) are:

$$\theta_X = \frac{d_X}{f_5}$$
$$\theta_Y = \frac{d_Y}{f_5}$$
(2)

## 3) Point Spread Function (PSF) of the synthetic image

*Raw paraxial impulse response*

We consider the situation of Figure 1 and Figure 2 where the target plane is situated at a distance L from the waist plane (focal image plane of $L_5$) and is scanned by a defocused beam. Because the imaging system is linear (if the object is laterally translated, the raw acquired image is simply translated too), we only need to consider the case of the response $h_R(L,x,y)$ of a punctual object located on the beam axis for $\alpha_X$ and $\alpha_Y = 0$. Then the target coordinates are $(L\theta_X, L\theta_Y)$. This is the Point Spread Function (PSF) of the imaging system at a distance acquisition of L from the waist plane. For a more complex image $t(x,y)$, the signal given by the imaging system is simply $t(x,y) * h_R(L,x,y)$, where * is the convolution operator. The electric filed striking the target can be simply derived from the propagation of complex electric field over a distance L. $h_R(L,x,y)$ then corresponds to the square of this electric field because of the symmetry between emission and reinjection of electric field:

$$h_R(L,x,y) \propto \left( \exp\left(-\frac{x^2+y^2}{r^2(1+(\frac{L}{Z_R})^2)}\right) \exp\left(j2\pi \frac{(x-L\theta_x)^2+(y-L\theta_y)^2}{2R(L)\lambda}\right) \right)^2$$

$$\propto \exp\left(-\frac{x^2+y^2}{RES_R(L)^2}\right) \exp\left(j2\pi \frac{(x-L\theta_x)^2+(y-L\theta_y)^2}{2\frac{R(L)}{2}\lambda}\right) \quad (3)$$

With :

$$RES_R(L) = \frac{r}{\sqrt{2}}\sqrt{1+(\frac{L}{Z_R})^2}$$

$$Z_R = \frac{\pi r^2}{\lambda} \quad (4)$$

$$R(L) = L\left(1+\left(\frac{Z_R}{L}\right)^2\right)$$

The square caused by symmetrical coupling between the object and the laser mode was forgotten in [20,21,22] an so an error of √2 was made on the resolution in these articles; we have corrected this error here. In these equations and in future equations, we only give proportional expressions because only the shape (related to resolution) is interesting. Considering far-field situation (L >> $Z_R$), $h_R(L,x,y)$ corresponds to a wavefront of lateral spatial width $RES_R(L)$ and a radius of curvature $R(L) / 2 \approx L / 2$. It is interesting to see the spectral content of the signal. Using the stationary phase theorem [25] on Eq. (3), we get $H_R(L,\upsilon,\mu)$ the Fourier transform of $h_R(L,x,y)$, where $(\upsilon,\mu)$ are the spatial frequencies associated with (x,y):

$$H_R(L,\nu,\mu) \propto \exp(-\frac{(\nu - F_X)^2}{\Delta \nu^2} + \frac{(\mu - F_Y)^2}{\Delta \mu^2}) \exp(-j\frac{\pi L \lambda ((\nu - F_X)^2 + (\mu - F_Y)^2)}{2}) \quad (5)$$

With:

$$F_X = \frac{2\theta_X}{\lambda}$$
$$F_Y = \frac{2\theta_Y}{\lambda} \quad (6)$$
$$\Delta \nu = \Delta \mu = \frac{\sqrt{2}}{\pi r}$$

Here too, it is important to give a more physical interpretation: each plane wave in the signal corresponds to a different spatial Doppler frequency. As a result $H_R(L,\upsilon,\mu)$ represents the decomposition of the signal into plane waves (each couple $\upsilon,\mu$ represents a plane wave). The first exponential term represents the plane waves content of the laser beam of widths $\Delta\upsilon$ and $\Delta\mu$ inversely proportional to the waist r and centered around central Doppler shift ($F_X$ and $F_Y$ proportional to $\theta_X$ and $\theta_Y$ as seen in Eq. (2)). The second exponential term (phase term) gives the phase shifts between these plane waves which are characteristic of the defocus of the laser beam. More precisely, this second term corresponds to the free space transfer function over a distance L / 2 as expected.

## Numerical refocusing process

It is possible to numerically refocus on the object by eliminating the quadratic dephasing between plane waves expressed in the second exponential term of Eq. (5). This is equivalent to multiply $H_R(L,\nu,\mu)$ by the free space transfer function over a distance – L / 2 (retropropagation) $H_{filter}(L,\nu,\mu)$ given by:

$$H_{filter}(L,\nu,\mu) = \exp\left( j\frac{\pi L \lambda (\nu^2 + \mu^2)}{2} \right) \tag{7}$$

After numerical filtering, we get a signal $H_{SA}(L,\nu,\mu)$ in the spatial frequency domain:

$$\begin{aligned} H_{SA}(L,\nu,\mu) &= H_R(L,\nu,\mu) H_{filter}(L,\nu,\mu) \\ &\propto \exp\left( -\frac{(\nu - F_x)^2 + (\mu - F_y)^2}{\Delta \nu^2} \right) \exp( j\pi \lambda L(\nu F_X + \mu F_Y)) \end{aligned} \tag{8}$$

If we calculate the inverse Fourier transform of $H_{SA}(L,\nu,\mu)$ we obtain $h_{SA}(L,x,y)$ the final numerically refocused image in the plane of the target:

$$\begin{aligned} |h_{SA}(L,x,y)| &= |TF^{-1}(H_{SA})(x,y)| \\ &\propto \left| \exp( -\pi^2 \Delta\nu^2 ((x + L\theta_x)^2 + (y + L\theta_y)^2)) \exp( j2\pi(xF_x + yF_y)) \right| \\ &\propto \exp\left( -\frac{((x + L\theta_x)^2 + (y + L\theta_y)^2)}{\left(\frac{r}{\sqrt{2}}\right)^2} \right) \end{aligned} \tag{9}$$

The final resolution is thus $r / \sqrt{2}$ which implies by comparing to Eq. (4) ($RES_R(0) = r / \sqrt{2}$) that the lens $L_5$ resolution is recovered despite defocused raw acquisition. It is important to note that after filtering, the resolution does not depend on L but only on the plane wave content of the signal and so on r.

*Final image resolution and comparison with raw imaging*

It has been previously shown that by numerically refocusing the raw image on the target plane, we get the same resolution $RES_R(0)$ as if we used a direct focusing on the object. The question we can now ask is what happens if we have a second object (object 2) in a plane at a distance $\delta$ from object 1 (see **Erreur ! Source du renvoi introuvable.** a)) and if we perform a numerical refocusing in the plane of object 1. What is the signal of object 2? Considering this, the new synthetic signal depends now on an additional parameter and is noted $h_{SA}(L-\delta,L,x,y)$ where the first parameter is the distance between the laser waist and the plane of object 2 (here $L-\delta$) and the second parameter the distance between the laser waist and the plane of refocusing containing objet 1 (here L). In the Fourier space, from the raw signal $H_R(L-\delta,\nu,\mu)$ (see Eq. (5)), we get after filtering by $H_{filter}(L,\nu,\mu)$ (see Eq. (8)) the final signal $H_{SA}(L-\delta,L,\nu,\mu)$ (the Fourier transform of $h_{SA}(L-\delta,L,x,y)$):

$$H_{SA}(L-\delta,L,\nu,\mu) = H_R(L-\delta,\nu,\mu) H_{filter}(L,\nu,\mu) \qquad (10)$$

To avoid unnecessary and complicated calculations, it is more convenient to have a physical vision to get the expression of $h_{SA}(L-\delta,L,x,y)$. We remind that $h_R(L-\delta,x,y)$ corresponds to the expression of a wavefront which have been propagated from a waist $r / \sqrt{2}$ on a distance $(L - \delta) / 2$ (width $RES_R(L-\delta)$), with angles ($2\theta_X$, $2\theta_Y$ / Z direction). As $H_{filter}$ is the free space transfer function over a distance $-L / 2$ (Eq. (7)), we immediately deduce:

$$|h_{SA}(L-\delta,L,x,y)| = |TF^{-1}(H_{SA}(L-\delta,L,\nu,\mu))(x,y)|$$
$$\propto \exp\left(-\frac{(x+L\theta_x)^2 + (y+L\theta_y)^2}{RES_{SA}(\delta)}\right) \quad (11)$$

with:

$$RES_{SA}(\delta) = r'\sqrt{1 + \left(\frac{\delta}{2Z_R'}\right)^2}$$
$$r' = \frac{r}{\sqrt{2}} \quad (12)$$
$$Z_R' = \frac{\pi r'^2}{\lambda}$$

$RES_{SA}(\delta)$ is the resolution of the signal $h_{SA}(L-\delta,L,x,y)$. By comparing Eq. (4) and Eq. (12) we note that $RES_{SA}(\delta) = RES_R(\delta)$. This implies that, considering the resolution, the numerical filtering by $H_{filter}(L,\nu,\mu)$ transforms the system into an equivalent one where lens $L_5$ is turned into a lens $L_{5eq}$ with a longer focal length $f_{5eq} = f_5 + L$ but keeping the same Numerical Aperture (NA) (see **Erreur ! Source du renvoi introuvable.**). As a result it is possible to get a higher working distance keeping a constant resolution and depth of field.

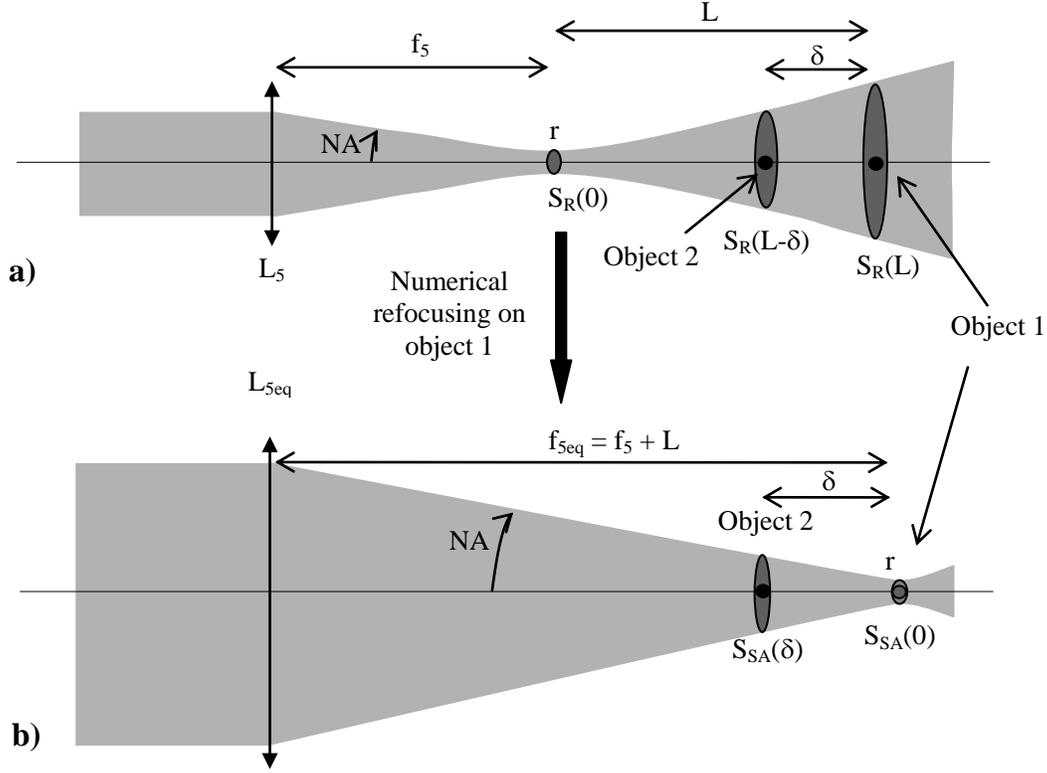

**Figure 3 : Comparison in terms of resolution of the a) : Raw acquisition and b) : after numerical refocusing in the plane of object 1. δ is the distance between two objects of interest. $S_R(L) = \pi \, RES_R(L)^2$ is the surface of the beam at a distance L from the beam waist and $S_{SA}(\delta) = \pi \, RES_{SA}(\delta)^2$ is the surface of the beam at a distance δ from the plane of refocusing (here plane of object 1).**

In addition to the capability of extending the working distance, this arrangement opens the possibility of fast 3D imaging since we are able to choose the plane of focusing numerically with a single X-Y object scanning. Further work about 3D imaging is planned in a close future.

Another thing to consider is the evolution of acquisition time and amount of data with the initial defocusing L. The number of data we have to store corresponds to the number of pixels in the raw image. The number of pixels needed corresponds to a correct sampling of the signal in the space – spatial frequency domain (the whole object field at Shannon frequency). In the case of the acquisition without defocusing (L = 0), since each raw pixel is acquired during an integration time T, the total acquisition $T_{acq}$ (L = 0) is:

$$T_{acq,\,R}(L=0) \propto L_X L_Y \Delta\nu\Delta\mu T \propto \frac{L_X L_Y T}{r^2} \tag{13}$$

In this expression $L_X$ and $L_Y$ are respectively the length and width of the target. In the case where $L \neq 0$, the raw initial defocus increases the spatial width of the signal by $2RES_R(L)$ in both X and Y directions; the spectral content stays unchanged (and so the sampling rate). As a result the total acquisition time is changed into $T_{acq}(L)$:

$$T_{acq}(L) \propto (L_X + 2RES_R(L))(L_Y + 2RES_R(L))\Delta\nu\Delta\mu T \propto \frac{(L_X + 2RES_R(L))(L_Y + 2RES_R(L))T}{r^2} \quad (14)$$

As a conclusion, the ability of going beyond the working distance of $L_5$ while keeping the resolution is possible at the cost of an increase of both time acquisition and quantity of data to store. This can be compared to angular galvanometric scanning developed in [22], where the time measurement stays constant whereas the resolution degrades with working distance.

Until that point, calculations have been performed for paraxial rays but if we consider all physical interpretations in term of plane wave phase manipulations, we can easily understand that it is possible to extend the analysis and the refocusing to lenses $L_5$ with higher numerical aperture (microscope objectives for example) or with spherical aberration. Instead of using paraxial filter $H_{filter}$ we need to use a filter which eliminates the phase shifts between the plane waves in the raw signal (to eliminate the second exponential of Eq. (5)), this filter can be calculated exactly if we know the aberrations of the lens or objective $L_5$.

## *Experimental validation*

The setup described on Figure 1 has been built and tested. We then have chosen a target made of reflective silica beads of 40 µm diameter behind a circular aperture of 1 mm diameter (Figure 4 a). The raw acquisition performed at a distance L = 1 cm is given in Figure 4 b), this image is enlarged compared to the object because of the defocusing; Figure 4 c) gives the modulus of its Fourier transform and Figure 4 d) the numerically refocused image. We successfully pass from a defocused image with a resolution $RES_R(L = 1\ cm) = 180\ \mu m$ to a numerically refocused image with a resolution $RES_{SA}(0) = RES_R(0) = 9.5\ \mu m$. The optimal refocusing distance was determined by using the detection criteria described by Dubois *et al.* [0]. The acquisition time for this 512*512 pixels image was approximately one minute with an integration time of $T = 50\ \mu s$ by pixel.

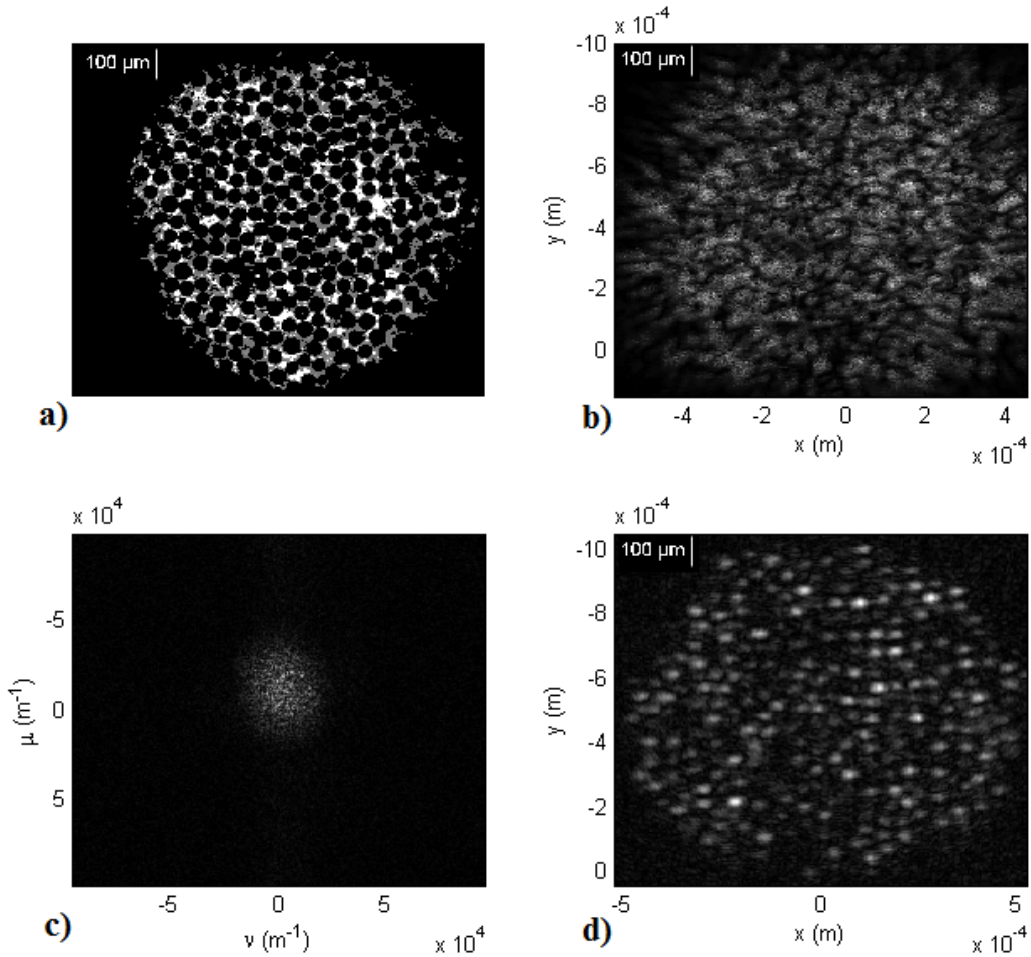

**Figure 4 : example of synthetic aperture LOFI. a) Object under microscope. It is made of reflective silica beads of 40 μm diameter behind a circular aperture of 1 mm diameter. The bright field transmission image is made through a Zeiss microscope objective with a magnification of 10 and a 0.25 numerical aperture (focal length of 20 mm). b) : Raw image of the object with LOFI setup r = 13.5 μm, L = 1 cm, 512*512 pixels. The beam size on the target plane is equal to 180 μm: beads are not resolved. c) : Recorded image in the space frequency domain. d) : Image after numerical refocusing, a resolution $RES_R(0) = RES_{SA}(0) = r / \sqrt{2} = 9.5\ \mu m$ is expected and verified experimentally.**

We now have to experimentally check the theoretical prediction of Eqs. (4) and (12) about the resolution of the raw and synthetic signals. In order to measure the resolution versus the defocus, we consider single silica bead (object Figure 4 a)) and we fit a section along the X direction with a Gaussian function. This method is possible because, relatively to the laser, a silica (spherical) bead behaves like a punctual reflector located at the geometric center of the bead. The different experimental resolutions versus the defocus (Figure 5) are then fitted with the theoretical predictions (RES$_R$(δ) and RES$_{SA}$(δ) in Figure 5 a) and Figure 5 b)) respectively.

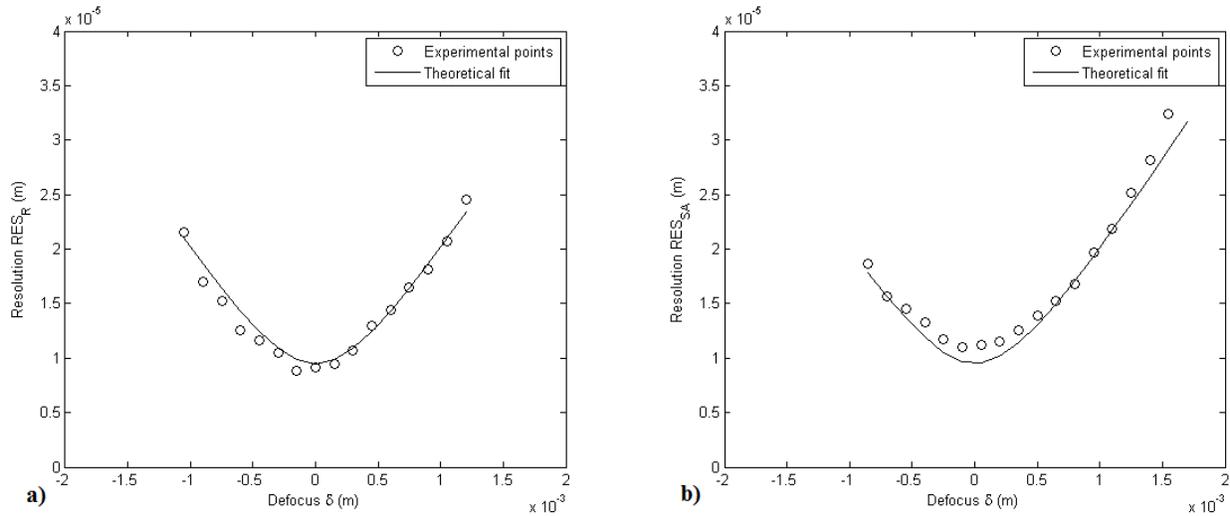

**Figure 5 : Evolution of the resolution in the X direction with the defocus δ, fitted by the theoretical expression of RES$_R$(δ). The resolution is calculated by fitting a section along X of the image of one bead in the image of Figure 4 (which is the PSF since the bead can be considered as a punctual scatterer). a) : Raw case, the defocus correspond to the L = δ in Figure 1 and Figure 2. b) Synthetic case, the defocus corresponds to the difference between the parameter L during the acquisition and the numerical retro propagation distance L' : δ = L – L' (see** Erreur ! Source du renvoi introuvable. **b)).**

We note that the theory is verified in both raw and synthetic cases. Fitting the experimental data leads to r / √2 ≈ 9.5 µm.

## 4) Photometric budget of the synthetic imaging setup

After the analysis of the resolution, we are now exploring photometric properties and more particularly the evolution of the signal power with the defocusing distance L (see Figure 1, Figure 2 and **Erreur ! Source du renvoi introuvable.**). First it is important to explain what we call "signal power": it corresponds to the number of photons reinjected in the laser and so which participate to the image. LOFI signal corresponds to an electric field and as a result, the signal power is given by the square of the modulus of the signal. Once again, we make the approximation of paraxial rays but we consider a uniform distribution of laser intensity instead of a Gaussian one to simplify the problem. We perform the calculations with a Lambertian diffusive target of albedo $\rho$ and of surface S. In order to have all the signal power concentrated in one pixel when we are focused on its plane (numerically or during the acquisition), the surface S is chosen smaller than $S_R(0) = \pi\ RES_R(0)^2$ (the object is not resolved). $S_R(L)$ is the surface of the beam at a distance L from the beam waist as shown in **Erreur ! Source du renvoi introuvable.** a). We first present the calculation for the raw acquisition with L = 0 already done in [22], then we make the analysis for L ≠ 0 and to conclude we experimentally verify the theoretical expectations.

### Direct focused LOFI

First we consider the case of raw acquisition directly in the target plane. The retrodiffused power $P_R$ in the only bright pixel of the image is given by [22]:

$$P_R = \rho L_R G_R = \rho \frac{E_R}{\pi} G_R = \frac{\rho P_0}{\pi S_R(0)} S \pi (NA)^2 = \frac{\rho P_0 S (NA)^2}{S_R(0)} \tag{15}$$

$L_R$ is the luminance and $E_R$ the illuminance in the focal image plane of $L_5$. $G_R$ is the geometrical extent, $P_0$ is the laser power sent on the object. $S_R(0)$ and the numerical aperture NA are given by :

$$S_R(0) = \pi RES_R(0)^2 = \frac{\pi r^2}{2}$$

$$NA = \left.\frac{RES_R(\delta)}{\delta}\right|_{\delta \gg Z_R} = \frac{\lambda}{\pi r \sqrt{2}} = \frac{\lambda}{2\sqrt{\pi S_R(0)}} \tag{16}$$

We finally rewrite Eq. (15) in this form:

$$P_R = \frac{\rho P_0 S (NA)^2}{S_R(0)} = \frac{\rho P_0 S \left(\frac{\lambda}{2\sqrt{\pi S_R(0)}}\right)^2}{S_R(0)} = \frac{\rho P_0 S \lambda^2}{4\pi S_R(0)^2} \tag{17}$$

## Synthetic LOFI

We now turn to the total power $P_{SA}(L)$ in the refocused image of a raw acquisition at a distance L $\neq$ 0. It is a bit more complicated since this image is built with several pixels $N_{pixels}(L)$ of the raw image due to the defocused scanning beam of surface $S_R(L) > S_R(0)$; the signal in each of the pixels of the raw image have a power $P_{RawSA}(L)$. After numerical refocusing the power from all these $N_{pixels}(L)$ pixels is concentrated in a single synthetic pixel. We have:

$$P_{SA}(L) = P_{RawSA}(L) N_{pixels}(L) \tag{18}$$

With:

$$P_{RawSA}(L) = \rho L_{SA}(L) G_{SA}(L) = \rho \frac{E_{SA}(L)}{\pi} G_{SA}(L) = \rho \frac{P_0}{\pi S_R(L)} \frac{SS_R(0)}{L^2} \tag{19}$$

$L_{SA}(L)$ is the luminance, $E_{SA}(L)$ the illuminance in the plane of object 1 (**Erreur ! Source du renvoi introuvable.** a)) and $G_{SA}(L)$ is the geometrical extent of the laser spot. The number of pixels in the raw image participating in the synthetic bright pixel is:

$$N_{pixels}(L) = \frac{S_R(L)}{S_R(0)} \qquad (20)$$

By combining Eq. (18), Eq. (19) and Eq. (20), we get:

$$P_{SA}(L) = P_{RawSA}(L) N_{pixels}(L) = \frac{\rho P_0}{\pi S_R(L)} \frac{SS_R(0)}{L^2} \frac{S_R(L)}{S_R(0)} = \frac{\rho P_0 S}{\pi L^2} \qquad (21)$$

Finally we expect a decrease of the signal power proportional to $L^2$. Making images deeper than working distance of lens $L_5$ has a cost in terms of signal power and then of Signal to Noise Ratio (SNR). However if we compare to [22], translational scanning shows a slower decrease of power (~1 / $L^2$) than rotational scanning (~ 1 / $L^4$). The quick decrease of $P_{RawSA}(L)$ with L ($\alpha$ 1/$L^4$) is linked to the fact that we use LOFI microscope outside its confocal zone; we can say that we have lost confocal advantage of LOFI. As a result in addition to the target signal attenuation, we have got another problem: multi-scattered photons are not rejected anymore compared to the signal of interest. However we have noticed that the increase of the background due to multi-scattered photons can be neglected compared to the attenuation of ballistic photons which is the true limitation for the accessible deepness through the scattering medium. We can think that it is due to low reflectivity of fatty milk particles (we use diluted milk as scattering medium).

## *Comparison between the two imaging modalities and discussion on the limitations*

We have shown the asymptotic dependence of $P_{SA}$ with L. We now predict a more quantitative relation between $P_{SA}$ and $P_R$. By definition:

$$S_R(L) = \pi RES_R(L)^2 = \pi \left(\frac{\lambda L}{\sqrt{2}\pi r}\right)^2 = \frac{\lambda^2 L^2}{4\pi \left(\frac{r}{\sqrt{2}}\right)^2} = \frac{\lambda^2 L^2}{4 S_R(0)}$$

$$\Rightarrow L = 2\frac{\sqrt{S_R(0) S_R(L)}}{\lambda} \quad (22)$$

By combining Eq. (17), Eq. (21) and Eq. (22), we finally get the following simple relation:

$$\frac{P_{SA}(L)}{P_R} = \frac{\dfrac{\rho P_0 S}{\pi L^2}}{\dfrac{\rho P_0 S \lambda^2}{4\pi S_R(0)^2}} = \frac{4 S_R(0)^2}{L^2 \lambda^2} = \frac{S_R(0)^2}{S_R(0) S_R(L)} = \frac{S_R(0)}{S_R(L)} \quad (23)$$

The ratio between two images acquired with and without defocusing is simply given by the ratio of surfaces of the beam in the two target planes of acquisition ($S_R(0)$ and $S_R(L)$). In the same manner than with the resolution (but now considering photometric balance) we can ask what the parasitic signal introduced by object 2 located at a distance L - δ from the waist will be, if we want to image object 1 (**Erreur ! Source du renvoi introuvable.**) considering that the two objects have the same albedo ρ and surface S. Similarly to the discussion on the resolution we need to introduce a new parameter in $P_{SA}(L)$ which become $P_{SA}(L-δ,L)$, the power from the object 2 in one pixel (which is perturbating the image of object 1) when we numerically refocus in the plane of object 1. The image of object 2 is then reduced to one pixel. We have the following relation:

$$\frac{P_{SA}(L,L)}{P_{SA}(L-\delta,L)} = \frac{\dfrac{P_{SA}(L,L)}{P_R}}{\dfrac{S_R(0)}{S_R(\delta)}\dfrac{P_{SA}(L-\delta,L-\delta)}{P_R}} = \frac{S_R(\delta)}{S_R(0)}\dfrac{\dfrac{S_R(0)}{S_R(L)}}{\dfrac{S_R(0)}{S_R(L-\delta)}} = \frac{S_R(\delta)}{S_R(0)}\dfrac{S_R(L-\delta)}{S_R(L)} \quad (24)$$

Because of the defocusing effect we have:

$$P_{SA}(L-\delta,L) = \frac{S_R(0)}{S_R(\delta)} P_{SA}(L-\delta,L-\delta) \quad (25)$$

In the last equality of Eq. (24), the first ratio corresponds to the defocus over a distance $\delta$ (**Erreur ! Source du renvoi introuvable.** b)) and the second factor of the final expression in Eq. (24) comes from Eq. (23) and is due to the total number of photons reinjected. It is important to remark that in the situation of a raw acquisition focused on object 2 (L - $\delta$ = 0), even if we refocus on object 1, the power of the parasitic signal in one pixel from object 2 in the synthetic image, is at the same level that the total power from object 1 (concentrated in only one pixel as we numerically refocus on it). As a result, in that situation, we are not able to separate object 2 from object 1 despite object 2 is defocused in the synthetic image; this is due to the excessive amount of power reinjected by object 2 during raw acquisition. Nevertheless, this situation is improved when both object 1 and 2 are acquired far from the waist (L is increased in **Erreur ! Source du renvoi introuvable.** a)). Indeed for high values of L (L >> $\delta$), Eq. (24) is simplified in $S_R(\delta) / S_R(0)$ (> 1). However we showed in Eq. (21) that the total power reinjected by an object located at a distance L is degrading proportionally to L². As a result in order to have both a good separation between object 1 and 2 and a globally sufficient SNR, we must find a compromise for L.

### *Experimental validation*

To confirm the theoretical predictions of the previous section, we have experimentally measured the power in the image of the previous object made with silica beads (Figure 4 a)). What we call the total power corresponds to the sum of the square modulus of the signal coming from all pixels of raw (Figure 4 b)) or refocused image (Figure 4 c)). Indeed, according to Parseval's theorem and to Eq. (10) where the filter transfer function has a modulus equal to unity, the power in raw and refocused image is the same. We have experimentally measured the power in the signal versus the defocus of the raw acquisition, these points have then been fitted with the theoretical curve given by Eq. (3) and (11). These results are presented in Figure 6.

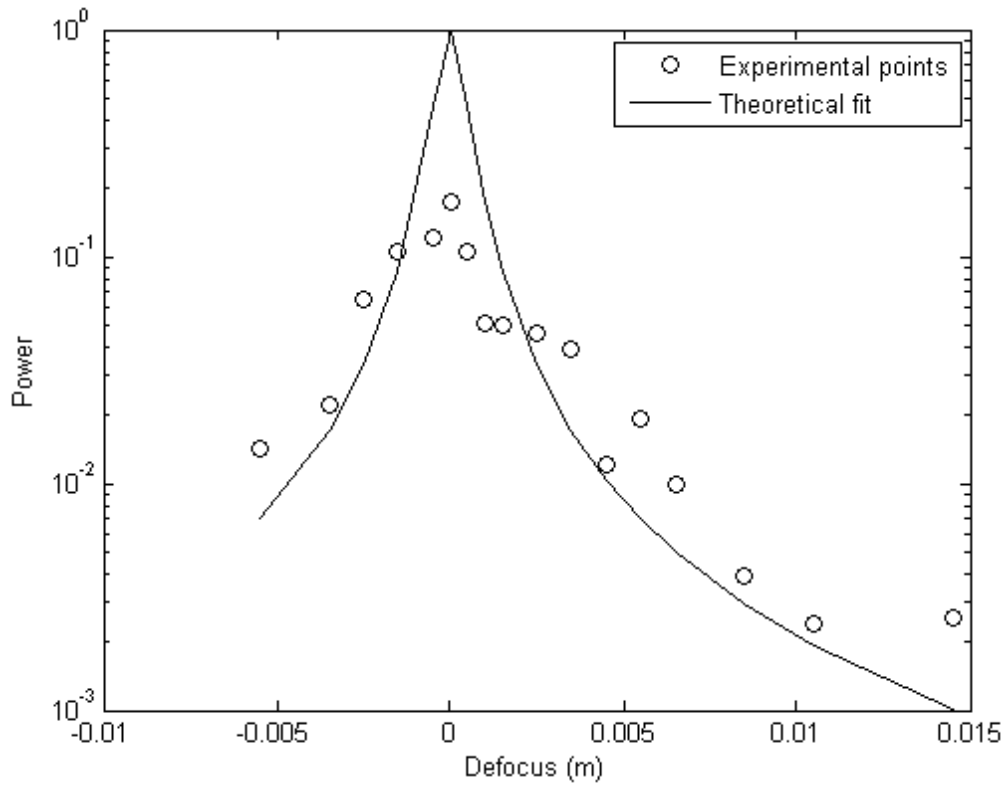

**Figure 6 : Evolution of the signal power reflected by the object of Figure 4 a) versus the defocus L (see Erreur ! Source du renvoi introuvable. a)). Positive defocus corresponds to an increase of the distance between the object and the lens $L_5$. The vertical axis has an arbitrary normalized unit.**

We can consider that our theoretical predictions describe relatively well the experimental results. The main discrepancy arises for small defocus values, where there is a difference of a factor ~ 10. This could be explained by several approximations that have been made: we have neglected the small astigmatism of the laser beam that is important for low values of L and the non-lambertian nature of silica beads in the target.

## 5) Conclusion and future work

We have made the demonstration of a simple, self-aligned and very sensitive imaging system able to work beyond the working distance of an objective lens, which is particularly useful to make resolved image through deep turbid media. Theses performances have been reached by combining the use of the LOFI technology which gives the sensitivity and the self-alignment, with the original 2D translational scanning associated to aperture synthesis which keeps the resolution constant beyond the usual lens working distance. Thank to our simple 2D translational scanning with galvanometric mirrors, we realized, to our knowledge, the first imaging system based on synthetic aperture numerical treatment with a fixed target and an isotropic resolution. Concerning the resolution, we have shown that performing a numerical refocusing enables to transform the initial real objective or lens into an equivalent objective with a higher focal length (thus a longer working distance) while keeping the initial numerical aperture (and so the resolution). But we demonstrated also that this advantage is at the cost of an increased acquisition time and data storage (proportional to L²) and reduced photometric performances (proportional to 1 / L²). We have characterized the 3D PSF of our synthetic microscope with 2D images (silica beads). In a future work, we plan to investigate real 3D imaging. Regarding the photometric performances, we explored the signal power reinjected versus the acquisition defocus and the influence of diffusers in different planes when we are numerically focusing in another plane. We also have to present the influence of all noises and perturbations on our final image in order to get the final accessible SNR; this will be presented in a companion paper.